\documentclass[twocolumn]{aastex62}
\usepackage{amsmath}
\usepackage{wasysym}
\usepackage{graphicx}
\usepackage{amssymb}
\usepackage{epstopdf}
\usepackage{mathrsfs}
\usepackage{anyfontsize}
\usepackage{natbib}
\usepackage{color}
\usepackage{lipsum}
\usepackage{diagbox}
\usepackage{gensymb}
\usepackage{booktabs}
\usepackage{textcomp}
\usepackage[T1]{fontenc}

\shorttitle{The effects of large-scale magnetic fields for CL AGN}
\shortauthors{Pan et al.}
\begin{document}
\title{The effects of large-scale magnetic fields on the model for repeating changing-look AGNs}
\correspondingauthor{Xin Pan, Shuang-Liang Li, Xinwu Cao}
\email{panxin@shao.ac.cn, lisl@shao.ac.cn, xwcao@zju.edu.cn}

\author[0000-0002-6938-3594]{Xin Pan}
\affiliation{Key Laboratory for Research in Galaxies and Cosmology, Shanghai Astronomical Observatory, Chinese Academy of Sciences, 80 Nandan Road, Shanghai 200030, China}
\affiliation{University of Chinese Academy of Sciences, 19A Yuquan Road, 100049, Beijing, China}

\author[0000-0002-7299-4513]{Shuang-Liang Li}
\affiliation{Key Laboratory for Research in Galaxies and Cosmology, Shanghai Astronomical Observatory, Chinese Academy of Sciences, 80 Nandan Road, Shanghai 200030, China}
\affiliation{University of Chinese Academy of Sciences, 19A Yuquan Road, 100049, Beijing, China}

\author[0000-0002-2355-3498]{Xinwu Cao}
\affiliation{Zhejiang Institute of Modern Physics, Department of Physics, Zhejiang University, 38 Zheda Road, Hangzhou 310027, China}
\affiliation{Shanghai Astronomical Observatory, Chinese Academy of Sciences, 80 Nandan Road,  Shanghai, 200030, China}

\begin{abstract}

Periodic outbursts are observed in several changing-look (CL) active galactic nuclei (AGNs). \citet{sniegowska_possible_2020} suggested a model to explain the repeating CL in these AGNs, where the periodic outbursts are triggered in a narrow unstable zone between an inner ADAF and outer thin disk. In this work, we intend to investigate the effects of large-scale magnetic fields on the limit cycle behaviors of CL AGNs. The winds driven by magnetic fields can significantly change the structure of thin disk by taking away the angular momentum and energy of the disk. It is found that the period of outburst in repeating CL AGNs can be substantially reduced by the magnetic fields. Conversely, if we keep the period unchanged, the outburst intensity can be raised for several times. These results can help to explain the observational properties of multiple CL AGNs. Besides the magnetic fields, the effects of transition radius $R_{\rm tr}$, the width of transition zone $\Delta R$ and Shakura-Sunyaev parameter $\alpha$ are also explored in this work.

\end{abstract}


\section{Introduction}

Active galactic nuclei (AGNs) can be classified into type 1 and type 2 based on the appearance or disappearance of broad emission lines \citep{seyfert_nuclear_1943}, which is usually explained as whether broad emission line region (BLR) is obscured by torus or not in AGN unified model (e.g., \citealt{antonucci_unified_1993}). Some AGNs have been reported that their types change on timescale of months to years for many years, i.e., from type 1 to 2 (e.g., \citealt{penston_evolutionary_1984,elitzur_evolution_2014}), or from type 2 to 1 (e.g., \citealt{khachikian_new_1971,katebi_ps1-13cbe_2019}), or even experience multiple changes (e.g., \citealt{mcelroy_close_2016,wang_suspended_2020}). These sources are the so-call changing-look AGNs (CL AGNs) and have attracted more and more attention in recent years. Up to 100 CL AGNs have been discovered so far (e.g. \citealt{parker_detection_2016,gezari_iptf_2017,yang_discovery_2018,macleod_changing-look_2019,wang_x-ray_2020}).

The physical origin of CL AGNs is still under debate. The first direct explanation is that BLR is obscured by clouds crossing over our line of sight (e.g., \citealt{goodrich_spectropolarimetry_1989,tran_extreme_1992,elitzur_unification_2012}). However, both the low polarization level and strong variability of infrared (IR) emissions in CL AGNs don't support this scenario (\citealt{macleod_systematic_2016, 2017ApJ...846L...7Sheng, hutsemekers_polarization_2019}). Tidal disruption events (TDEs) is also a possible mechanism for some CL AGNs (\citealt{merloni_tidal_2015,kawamuro_hard_2016,ricci_destruction_2020}), but other CL AGNs may be not triggered by TDEs (e.g., \citealt{alloin_recurrent_1986, 2016MNRAS.455.1691Runnoe, wang_suspended_2020}). Besides these two mechanisms, change of mass accretion rate seems to be the most promising candidate for CL AGNs as suggested by recent works (e.g. \citealt{elitzur_evolution_2014,ross_new_2018,2019arXiv191203972L,sniegowska_possible_2020}). However, the viscous timescale of a thin disk corresponding with variable accretion rate is found to be inconsistent with observed timescale of CL AGNs (e.g. \citealt{gezari_iptf_2017, 2018ApJ...864...27Stern}). This problem can be qualitatively resolved when taking the large-scale magnetic field into account \citep{2019MNRAS.483L..17Dexter}, or considering the instability of accretion disk \citep{sniegowska_possible_2020}. Lastly, the close binaries of supermassive black holes with high eccentricities has also been suggested to be a possible model for CL AGNs recently \citep{wang_changing-look_2020}.

For CL AGNs with periodic or quasi-periodic outbursts, the outburst mechanism should be related to some kind of instability. The inner region of a thin disk, which is dominated by radiation pressure, is known to be both thermally and viscously unstable \citep{shakura_black_1973, 1976MNRAS.175..613Shakura}, unless other factors are included, such as, convection, turbulence, or magnetic field \citep{1995ApJ...443..187G,2011ApJ...732...52Z,li_thermal_2014,2015ApJ...801...47Y}. \citet{sniegowska_possible_2020} proposed an ingenious model to explain the repeating CL AGNs. In order to decrease the viscous timescale, the instable region is limited to a narrow zone between outer thin disk dominated by gas pressure and inner advection-dominated accretion flows (ADAF). This toy model can qualitatively repeat the observed multiple outbursts. In observations, relativistic jets are discovered in many AGNs (e.g., see a review in \citealt{2019ARA&A..57..467Blandford} and references therein), where large-scale magnetic fields play a key role in the popular mechanism launching relativistic jets \citep{blandford_electromagnetic_1977, hawley_disks_2015}. Winds are also found to be present in a significant fraction of AGNs through the observation of blue-shift absorption/emission lines (e.g., \citealt{2020MNRAS.492.5540Matthews}). Except for radiation pressure and line-driven (e.g., \citealt{2000ApJ...543..686Proga,2004ApJ...616..688Proga}), large-scale magnetic fields can also drive disk winds in some AGNs (e.g., \citealt{blandford_hydromagnetic_1982, 1986A&A...156..137Camenzind, 2014ApJ...783...51Cao}). The magnetic accelerated winds can significantly change the structure of disk by taking away both the angular momentum and energy. The former will decrease disk temperature and the latter can improve radial velocity of disk \citep{li_thermal_2014}. Therefore, we anticipate that large-scale magnetic fields can affect both the period and outburst amplitude of disk instability. In this work, we will investigate the effects of magnetic fields on the model produced by \citet{sniegowska_possible_2020} and on repeating CL AGNs.


\section{Model}
\label{sec:model}

\begin{figure}[htbp]
   \centering
   \includegraphics[width=0.47\textwidth]{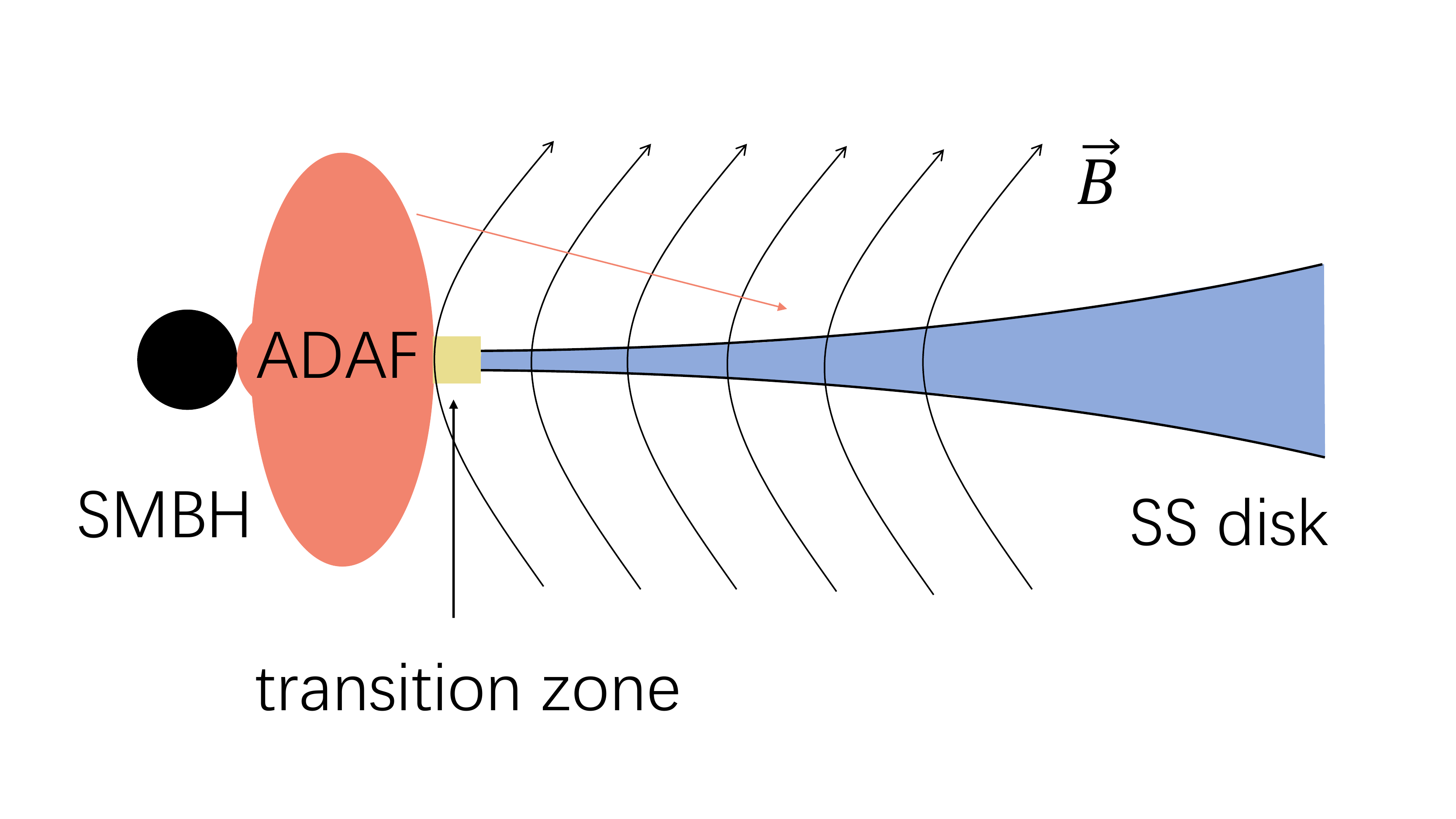}
   \caption{The schematic picture of our model. The intermediate zone is unstable, locating between an inner ADAF and outer gas-dominated thin disk, where the irradiation from inner ADAF is included.}
   \label{fig:sch}
\end{figure}

The inner region of a thin disk is known to be unstable \citep{shakura_black_1973, 1976MNRAS.175..613Shakura}. Based on this phenomenon, \citet{sniegowska_possible_2020} proposed a possible mechanism for the multiple CL AGNs, where they assumed that the accretion flow is composed of an inner ADAF and outer thin disk. The radiation-dominated region of a thin disk will shrink when mass accretion rate decreases, in the mean while an inner ADAF will appear. A specific mass accretion rate $\dot{m}_{\rm st}$ may be present, for which the inner ADAF will coincide with the outer radius of instable region in a thin disk (see \citealt{sniegowska_possible_2020} for details). When the accretion rate $\dot{m}$ is slightly higher than $\dot{m}_{\rm st}$, a small instable belt ($\Delta R$) will emerge between outer gas-dominated disk and inner ADAF. Because the belt is very narrow ($\Delta R \ll R$), its viscous timescale can be decreased to match the observed timescale of CL AGNs. The irradiation of inner ADAF on outer thin disk is also included in \citep{sniegowska_possible_2020}, where the radiative efficiency is simply adopted to be 10 percent \citep{bisnovatyi-kogan_influence_1997,ferreira_jet_2011,hirotani_very_2018}.  In this work, we study the effects of magnetic fields on the model suggested by \citet{sniegowska_possible_2020}. As shown in figure \ref{fig:sch}, large-scale magnetic fields are assumed to thread not only on the radiation-dominated belt but also on the outer gas-dominated disk.

\subsection{Model for outer thin disk dominated by gas pressure}

At first, we present the equations for outer thin disk region threaded by large-scale magnetic fields. In all the calculations, we adopt pseudo-Newtonian potential \citep{1980A&A....88...23Paczynsky} and $\alpha P$ prescription for viscosity. Following \citet{li_thermal_2014}, the continuity equation is given by
\begin{equation}
    \frac{d\dot{M}}{dR}+4\pi R\dot{m}_{\rm w}=0,
    \label{continuity}
\end{equation}
where $\dot{m}_{\rm w}$ is the mass loss rate from unit surface area of disk.  $\dot{m}_{\rm w}$ can be gotten from 
\begin{equation}
    \dot{m}_{\rm w}=\frac{B_{\rm p}B_{\rm z}}{4\pi\Omega_{\rm k} R}\mu,
    \label{massloss}
\end{equation}
where $\mu$ is the dimensionless, mass loading parameter of the outflow \citep{2013ApJ...765..149Cao}, in this work we adopt $\mu=0.001$ for fast moving winds with low mass-loss rates \citep{1998ApJ...499..329Ogilvie, 2002MNRAS.332..999Cao}. Here $B_{\rm p}$ ($=\sqrt{B_{\rm z}^2+B_{\rm r}^2}$) and $B_{\rm z}$ are the poloidal and vertical component of magnetic fields, respectively. In order to successfully launch jets from a cold disk, the inclination angle of field line with respect to disk surface should be smaller than $60^\circ$  \citep{blandford_hydromagnetic_1982}. We simply adopt an angle of $60^\circ$ in this work. 

The angular momentum equation reads
\begin{equation}
    -\frac{\dot{M}}{2\pi}\frac{d(\Omega_{\rm k} R^2)}{dR}+\frac{d}{dR}(2\alpha PHR^2)+T_{\rm m} R=0,
    \label{angularmom}
\end{equation}
where $T_{\rm m}$ is the magnetic torque exerted on accretion disk. The magnetic torque exerted on the accretion flow is \citep{li_thermal_2014}
\begin{equation}
    T_{\rm m}=\frac{B_{\rm p}B_{\varphi}R}{2\pi},
    \label{magtorque1}
\end{equation}
which can also be given through the outflows \citep{cao_large-scale_2013}
\begin{equation}
    T_{\rm m}=\frac{3}{4\pi}RB_{\rm p}^{2}\mu\left(1+\mu^{-2/3}\right).
    \label{magtorque2}
\end{equation}
Combining the equation (\ref{magtorque1}) and (\ref{magtorque2}) we can derive
\begin{equation}
    \frac{B_{\varphi}}{B_{\rm p}}=\frac{3}{2}\mu\left(1+\mu^{-2/3}\right),
\end{equation}
in the case of $\mu=0.001$, the toroidal component of magnetic field $B_{\varphi}=0.15B_{\rm p}$ . The total pressure of accretion disk is given as
\begin{equation}
    P=(1+\frac{1}{\beta_1})(P_{\rm gas}+P_{\rm rad}),
    \label{EoS}
\end{equation}
where $\beta_1=(P_{\rm gas}+P_{\rm rad})/(B^2/8\pi)$. The gas pressure $P_{\rm gas}$ and radiation pressure $P_{\rm rad}$ are given by $P_{\rm gas}=2\rho k_{\rm B} T_{\rm c}/m_{\rm H}$ and $P_{\rm rad}=aT_{\rm c}^4/3$, repectviely.

The energy equation can be calculated with
\begin{equation}
\nu\Sigma R^2 \left(\frac{d\Omega_{\rm k}}{dR}\right)^2+\frac{L_\star(1-a_\star)(H_\star+q_\star H)}{4\pi\left[R^2+(H_\star-H)^2\right]^{3/2}}=\frac{8acT_c^4}{3\tau}
    \label{energy}
\end{equation}
where $\nu$ is the viscosity coefficient and $\Sigma=2\rho H$ is the surface density. The optical depth is given by $\tau=\bar{\kappa}\Sigma/2$, where $\bar{\kappa}$ is the opacity. The second in the left side of equation represents the irradiation from inner ADAF. $L_\star, H_\star, a_\star$ and $q_\star$ correspond to the luminosity of irradiation, height of irradiation source, albedo and an index for $H/
R\propto R^{q_\star}$, respectively (see \citealt{liu_accretion_2016} for details). With above equations (\ref{continuity}), (\ref{angularmom}),  (\ref{EoS}) and (\ref{energy}), we can get steady solutions for outer thin disk.

\subsection{Model for transition zone dominated by radiation pressure}

The winds driven by magnetic fields can take away the mass, angular momentum and energy from accretion disk. Therefore, we need to revise the model produced by \citet{sniegowska_possible_2020}. Following their work, we assume that the inflow rate from outer thin disk to the transition zone is constant and that the evaporation rate from transition zone to inner ADAF is proportional to the scale-height and surface density of transition zone. This evaporation rate can be written as:
\begin{equation}
    \dot{M}=\dot{M}_0\frac{H}{H_0}\frac{\Sigma}{\Sigma_0},
    \label{evaporation}
\end{equation}
where $\dot{M}_0$, $H_0$ and $\Sigma_0$ are the inflow rate from outer zone to transition zone, scale-height and surface density of transition zone at equilibrium state, respectively. With $\dot{M}$, the time evolution equation of surface density can be revised as
\begin{equation}
    \frac{d\Sigma}{dt}=\frac{\dot{M}_0-\dot{M}-4\pi r\dot{m}_{\rm w} \Delta R}{2\pi R\Delta R},
    \label{denevolution}
\end{equation}
where $\Delta r$ is the width of transition zone.

The general form of time evolution equation for energy is 
\begin{equation}
    \Sigma T\frac{ds}{dt}=Q^+-Q^-,
    \label{energycons}
\end{equation}
where $Q^+$ and $Q^-$ are the viscous heating rate and radiative cooling rate, respectively. From the first law of thermodynamics, we have
\begin{equation}
    Tds=dq=\left(\frac{\partial u}{\partial T}\right)_\rho dT+\left[\left(\frac{\partial u}{\partial\rho}\right)_T-\frac{P}{\rho^2}\right]d\rho,
    \label{entropy}
\end{equation}
where $s$ is the entropy. Including energy of magnetic fields, the internal energy $u$ can be written as 
\begin{equation}
u=\frac{3P_{\rm gas}}{2\rho}+\frac{aT^4}{\rho}+\frac{B^2}{8\pi \rho}.
\end{equation}

Combining equation of state (\ref{EoS}), equation (\ref{entropy}) can be rewritten in a form similar to equation (13) in \citet{janiuk_radiation_2002}:
\begin{equation}
\begin{aligned}
    dq=&\frac{\beta_1}{1+\beta_1}\frac{P}{\rho}\left[\left(12-10.5\beta_2+\frac{4}{\beta_1}-\frac{3\beta_2}{\beta_1}\right)d\ln T\right.\\
    &\left.-\left(4-3\beta_2+\frac{1}{\beta_1}\right)d\ln\rho\right],
    \label{heatcontent3}
\end{aligned}
\end{equation}
where $\beta_2=P_{\rm gas}/(P_{\rm gas}+P_{\rm rad})$ is the ratio of gas pressure to the sum of gas pressure and radiation pressure. Finally, we can derive the temperature evolution equation as:
\begin{equation}
\begin{aligned}
    \frac{d\ln T}{dt}=&\frac{(Q^+-Q^--Q_{\rm adv})(1+\frac{1}{\beta_1})(1+\beta_2)}{2PH(28-22.5\beta_2-1.5\beta_2^2+\frac{8-2\beta_2-3\beta_2^2}{\beta_1})}\\
    &+2\frac{d\ln\Sigma}{dt}\frac{4-3\beta_2+\frac{1}{\beta_1}}{28-22.5\beta_2-1.5\beta_2^2+\frac{8-2\beta_2-3\beta_2^2}{\beta_1}}
    \label{temevolution}
\end{aligned}
\end{equation}
where $Q_{\rm adv}=\dot{M}PH/2\pi R \Delta R \Sigma$ is the advection cooling rate (see \citealt{sniegowska_possible_2020}).

\section{Numerical Results}

In this section, we start to investigate the effects of magnetic fields on the limit cycle behaviors of transition zone through equations (\ref{denevolution}) and (\ref{temevolution}). The albedo $a_\star=0.3$, $H_\star=10R_g$ ($R_{\rm g}=2GM/c^2$) and $q_\star=0.3$ are always adopted when calculating the irradiation from ADAF. Given the black hole mass $M$, the external accretion rate of transition zone $\dot{m}$ ($=\dot{M}/\dot{M}_{\rm Edd}$, $\dot{M}_{\rm Edd}=48\pi GMm_{\rm H}/\sigma_{\rm T}c^2$), the Shakura-Sunyaev parameter $\alpha$, the position of transition zone $R_{\rm tr}$, the width of transition zone $\Delta R$ and the parameter of magnetic field $\beta_1$, we can numerically solve the equations (\ref{denevolution}) and (\ref{temevolution}).

\begin{figure}[htbp]
   \centering
   \includegraphics[width=0.48\textwidth]{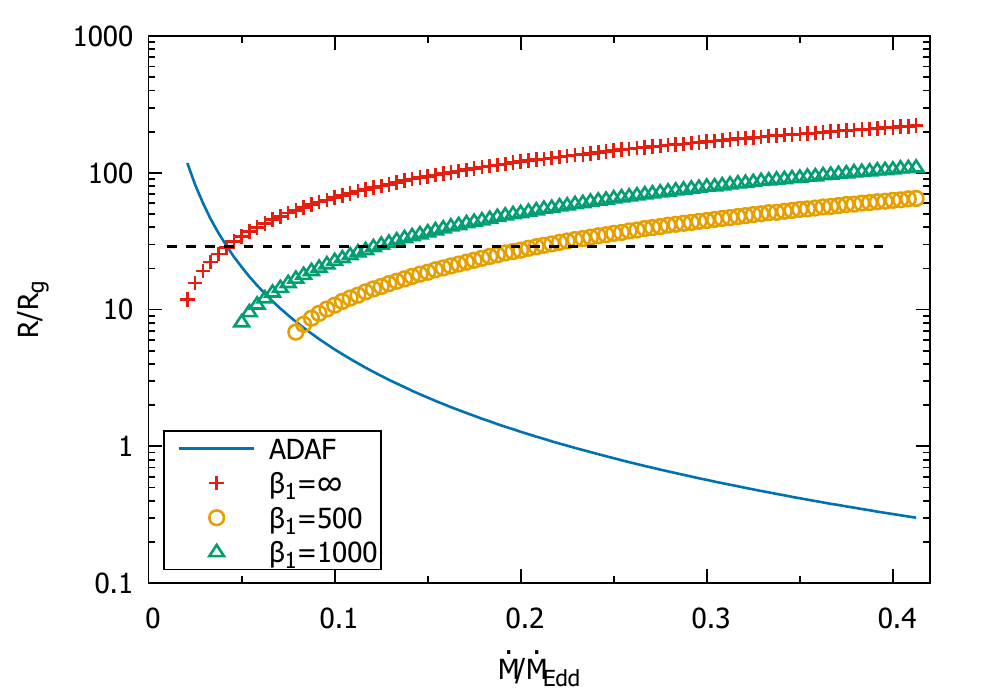}
   \caption{The outer radius of inner thin disk dominated by radiation pressure as functions of mass accretion rate, where the black hole mass $M=10^7 M_{\odot}$ and $\alpha=0.04$ are adopted. The blue line is the radius between ADAF and outer disk. 
   the red cross, green triangle and yellow circle represent $\beta_1=\infty, 1000, 500$, respectively.}
   \label{fig:trans}
\end{figure}

The winds driven by large-scale magnetic fields can reduce the temperature of disk by taking away its energy, which will affect the instability of disk \citep{li_thermal_2014}. In figure \ref{fig:trans}, we present that how the transition radius varies with the strength of magnetic field. The blue solid line shows the radius between an inner ADAF and outer thin disk, which is given by $R_{\rm ADAF}= 2 \alpha_{0.1}^{4} \dot{m}^{-2} R_{\rm g}$ in \citet{sniegowska_possible_2020}. It is found that the outer radius of inner thin disk dominated by radiation pressure decreases with increasing magnetic field strength, resulting on a smaller transition radius ($R_{\rm tr}$, the point of intersection between the blue line and other lines). However, this issue can be somewhat overcome when considering the effects of magnetic fields on $\alpha$. The value of $\alpha$ is found to increase with increasing magnetic field strength, leading to a higher $R_{\rm ADAF}$ (e.g., \citealt{2013ApJ...767...30Bai,2016MNRAS.457..857Salvesen}).

\begin{table}[htbp]
    \centering
    \caption{Detailed parameter of our calculations}
    \begin{tabular*}{0.47\textwidth}{ccccccc}
    \hline
    \hline
        Number & $\alpha$ & $\dot{m}$ & $R_{\rm tr}/R_{\rm g}$ & $\Delta R/R_{\rm g}$ & $\beta_1$ & $L_{\rm max}/L_{\rm min}$ \\
    \hline
        1 & 0.04 & 0.048 & 30 & 0.1 & $\infty$ & 7.93 \\
        2 & 0.04 & 0.240 & 30 & 0.1 & 500 & 7.17 \\
        3 & 0.04 & 0.138 & 30 & 0.1 & 1000 & 9.44 \\
        4 & 0.04 & 0.225 & 30 & 0.47 & 500 & 50.38 \\
        5 & 0.04 & 0.133 & 30 & 0.28 & 1000 & 31.71 \\
        6 & 0.04 & 0.065 & 40 & 0.1 & $\infty$ & 5.27 \\
        7 & 0.04 & 0.084 & 50 & 0.1 & $\infty$ & 3.77 \\
        8 & 0.06 & 0.045 & 30 & 0.1 & $\infty$ & 8.68 \\
        9 & 0.1 & 0.042 & 30 & 0.1 & $\infty$ & 9.84 \\
    \hline
    \end{tabular*}
    \label{tab:pars}
\end{table}

In order to investigate the effects of magnetic field and other parameters on the limit cycle behavior of transition zone, we provide nine group of parameters (see table \ref{tab:pars}). The parameters in group 1 are the default values in our calculations, while parameters in other group are just slightly modified. In all the calculations in table \ref{tab:pars}, the black hole mass $M=10^7 M_{\odot}$ is always adopted. 

\begin{figure}[htbp]
   \centering
   \includegraphics[width=0.5\textwidth]{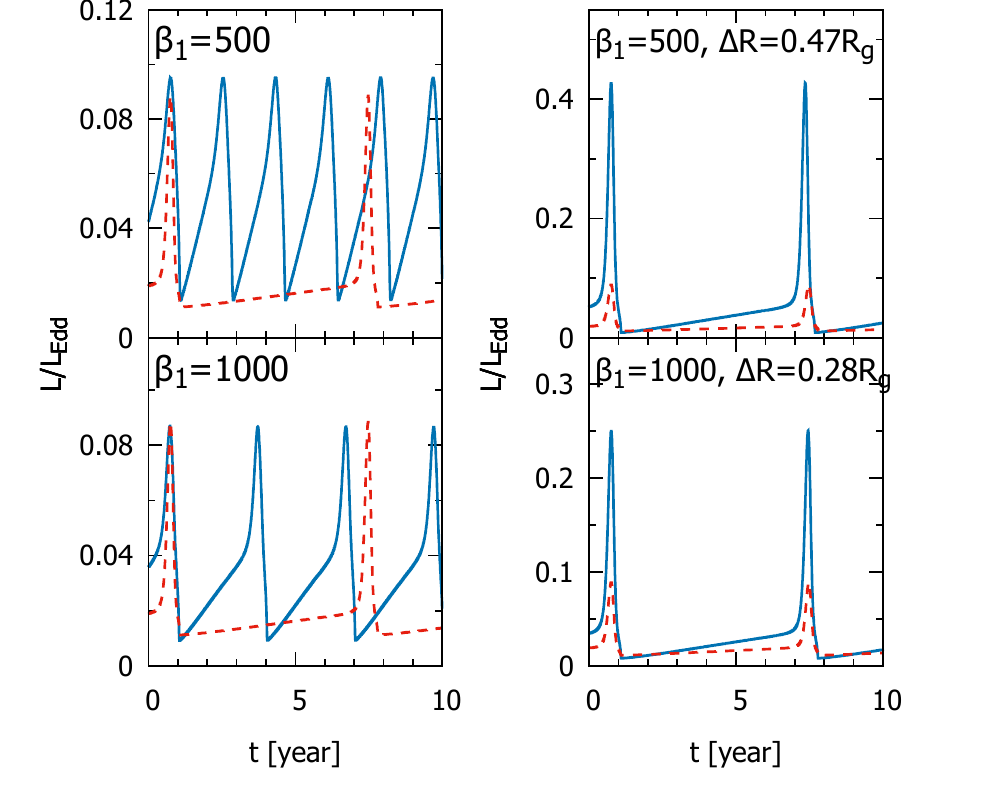}
   \caption{The light curves correspond to the parameters in group number 1, 2, 3, 4 and 5. The red dash line represents the light curve of the default parameters (group 1) and the blue solid lines represent the light curves of other group. Left panel: the comparison of light curves with different magnetic field strength (group 2 and 3). Right panel: the comparison of light curves with different $\Delta R$ and magnetic field strength (group 4 and 5).}
   \label{fig:comp1}
\end{figure}

\begin{figure}[htbp]
   \centering
   \includegraphics[width=0.5\textwidth]{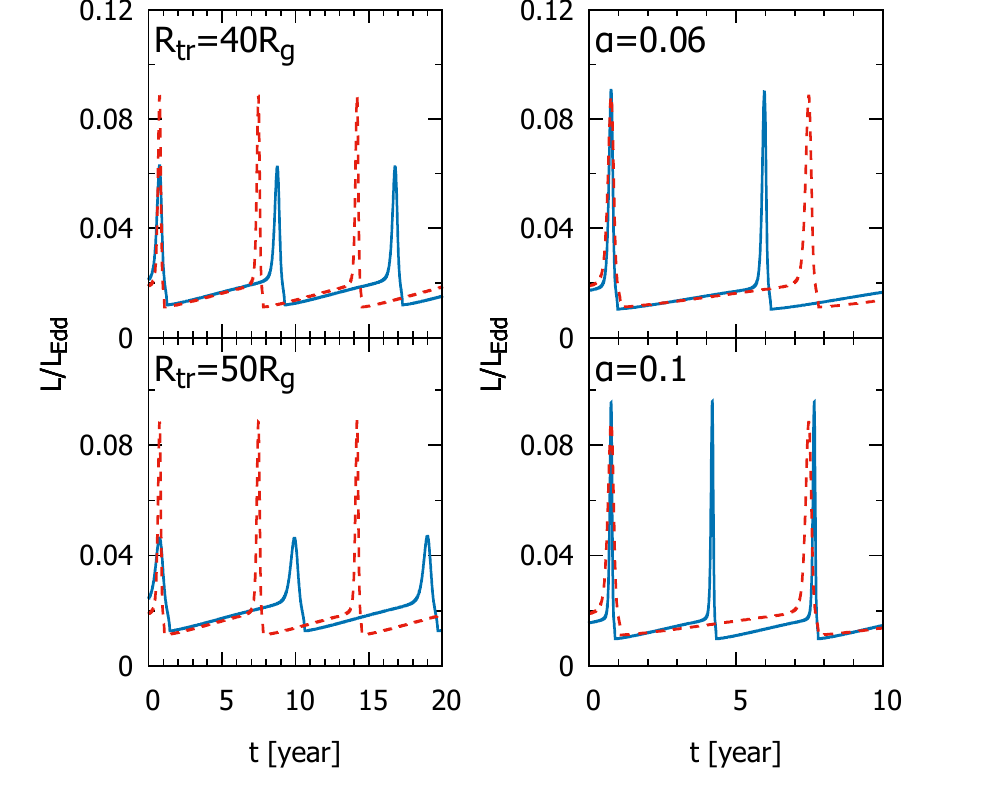}
   \caption{The light curves correspond to the parameters in group number of 1, 6, 7, 8 and 9. Left panel: the comparison of light curves with different $R_{\rm tr}$ (group 6 and 7). Right panel: the comparison light curves with different $\alpha$ (group 8 and 9).}
   \label{fig:comp2}
\end{figure}

The strength of magnetic fields ($\beta_{1}$) and the width of transition zone ($\Delta R$) are found to the limit cycle behaviors more significantly than other parameters. The period of disk light curve decreases fast with increasing strength of magnetic field (decreasing $\beta_{1}$) as shown in the left panel of figure \ref{fig:comp1}, while the shape of light curves are quite similar. If we keep the period of light curve as constant, $\Delta R$ is required to increase with decreasing $\beta_{1}$ (right panel of figure \ref{fig:comp1}). In this case, it is found that the outburst intensity ($L_{\rm max}/L_{\rm min}$) significantly increases with decreasing $\beta_{1}$. Notably, when tuning the parameter $\beta_{1}$, we just increase the mass accretion to ensure that the position of transition radius $R_{\rm tr}$ is roughly constant, while its width $\Delta R$ is assumed to remain unchanged for simply (group 2 and 3). The detailed calculation of $\Delta R$ should be carried out by considering the effects of large-scale magnetic fields on the heating and cooling process in ADAF, which is beyond the scope of this work.  

The position of transition zone $R_{\rm tr}$ and $\alpha$ can just slightly change the limit cycle behavior of transition zone (figure \ref{fig:comp2}). The increasing $R_{\rm tr}$ will lead to larger period and smaller outburst intensity simultaneously, as shown in the left panel of figure \ref{fig:comp2}. However, increasing $\alpha$ can only increase the period of light curve (right panel of figure \ref{fig:comp2}).

\section{Conclusions and Discussion}

In this work, we construct a new model to investigate the effects of large-scale magnetic fields on the model produced by \citet{sniegowska_possible_2020}. The presence of magnetic fields can greatly reduce the period of outburst in multiple CL AGNs. However, if the period is remained the same, the outburst intensity will increase for several times (see figure \ref{fig:comp1}). Besides magnetic fields, the width of transition zone ($\Delta R$), the position of transition zone $R_{\rm tr}$ and $\alpha$
can all change the limit cycle behaviors of transition zone. The large-scale magnetic fields adopted is very weak in our model. In case of strong magnetic fields, where MRI process will be suppressed \citep{2003PASJ...55L..69N}, reconnecting tearing instabilities may be responsible for the transport of angular momentum and produce the outburst in observations \citep{2010A&A...518A...5D,2011ApJ...743..192E}. 

The formation mechanism of large-scale magnetic fields in a standard thin disk is still an open issue so far. There are mainly two candidates currently. Firstly, the weak large-scale magnetic fields locating in outer region of a thin disk is difficult to be effectively dragged inward because the diffusive speed of magnetic field is found to be faster than its advection speed \citep{1989ASSL..156...99Van,1994MNRAS.267..235Lubow}. However, this problem may be solved when considering the effect of winds accelerated by magnetic fields. The strong winds driven by magnetic fields can greatly improve the advection speed by taking away most of angular momentum in a thin disk. Even for a very weak magnetic fields ($\beta_{1} > 100$), the initial magnetic fields can be effectively magnified \citep{2013ApJ...765..149Cao, li_thermal_2014}. Secondly, the dynamo process from magnetorotational instability (MRI) \citep{1991ApJ...376..214B} can also generate the large-scale magnetic fields, as suggested from both the shearing box and global simulations (e.g., \citealt{2015MNRAS.447...49S,2016MNRAS.459.1422E,2016MNRAS.462..818B}).

Our model can help to explain the observational properties of periodic repeating CL AGNs. For example, the fluxes in CL AGN NGC 1566 appear obvious periodicity \citep{alloin_recurrent_1986}. However, the outburst intensity of NGC 1566 is different in each outburst, which may be caused by the variation of magnetic field strength.  As shown in figure \ref{fig:comp1}, the outburst intensity can increase 6 times when considering a weak magnetic fields ($\beta_{1}=500$). GSN 069 is another CL AGNs showing periodic light curve, whose period is about 9 hours \citep{miniutti_nine-hour_2019}. In order to get such a short period, \citet{sniegowska_possible_2020} suggested that a small transition radius ($R_{\rm tr}$) and a big $\alpha$ are necessary. The presence of magnetic field can be in favor of shortening its period.  

\section* {ACKNOWLEDGEMENTS}

We thank the reviewer for helpful comments. SLL thanks Dr. Minfeng Gu and Weimin Gu for helpful discussion. This work is supported by the NSFC (grants 11773056, 11773050, 11833007, and 12073023).


\bibliography{ref}{}
\bibliographystyle{aasjournal}

\end{document}